\documentclass{IEEEtran}

\usepackage{amsmath}
\usepackage{amssymb}
\usepackage{flushend}
\usepackage{quantum}
\usepackage{color}
\usepackage[linktocpage,hypertexnames=false,pdfpagelabels]{hyperref}
\usepackage[amsmath,thmmarks,hyperref]{ntheorem}
\usepackage[capitalise,poorman]{cleveref}

\newtheorem{theorem}{Theorem}
\newtheorem{lemma}[theorem]{Lemma}
\newtheorem{corollary}[theorem]{Corollary}

\newtheorem{definition}[theorem]{Definition}

\theoremstyle{break}

\theoremstyle{break}
\theorembodyfont{\normalfont}

\theoremstyle{nonumberplain}
\theorembodyfont{\normalfont}
\theoremsymbol{$\Box$}

\DeclareMathAlphabet{\mathitbf}{OML}{cmm}{b}{it}

\newcommand{\gr}{\mathrm{Gr}}
\newcommand{\keyword}[1]{\emph{#1}}
\newcommand{\flip}{\mathbb{F}}
\DeclareMathOperator{\swap}{SWAP}
\newcommand{\mat}{\mathbb{M}}

\newcommand{\cE}{\mathcal{E}}
\newcommand{\cN}{\mathcal{N}}
\newcommand{\cR}{\mathcal{R}}
\newcommand{\cM}{\mathcal{M}}

\begin{document}

\title{An Extreme form of Superactivation\\for Quantum Zero-Error Capacities}

\author{Toby S.\ Cubitt and Graeme Smith

  \thanks{T.~S.~Cubitt is with the Departamento de An\'alisis
    Matem\'atico, Universidad Complutense de Madrid, Plaza de Ciencias~3,
    Ciudad Universitaria, 28040~Madrid, Spain (email:
    tcubitt@mat.ucm.es).}%
  \thanks{G.~Smith is with IBM T.J. Watson Research Center Yorktown
    Heights, NY 10598, USA (email: gbsmith@gmail.com)}%
  \thanks{T.~S.~Cubitt was supported by a Leverhulme early-career
    fellowship, and through the integrated EC project ``QAP'' (contract
    no.~IST-2005-15848). G.~Smith was supported by DARPA QUEST contract
    HR0011-09-C-0047. This work was carried out when T.~S.~Cubitt was
    with the Department of Mathematics, University of Bristol, United
    Kingdom}%
  \thanks{TSC would like to thank Nilanjana Datta and Francesco Buscemi
    for stimulating discussions about these ideas, and the physics of
    information group at IBM T.~J.~Watson Research Center for their
    hospitality during the visit in which this work was carried out.}}

%\date{\today}

\maketitle

\begin{abstract}
  The zero-error capacity of a channel is the rate at which it can send
  information perfectly, with zero probability of error, and has long
  been studied in classical information theory. We show that the
  zero-error capacity of quantum channels exhibits an extreme form of
  non-additivity, one which is not possible for classical channels, or
  even for the usual capacities of quantum channels. By combining
  probabilistic arguments with algebraic geometry, we prove that there
  exist channels $\cE_1$ and $\cE_2$ with no zero-error \emph{classical}
  capacity whatsoever, $C_0(\cE_1) = C_0(\cE_2) = 0$, but whose joint
  zero-error \emph{quantum} capacity is positive, $Q_0(\cE_1\ox \cE_2)
  \geq 1$. This striking effect is an extreme from of the
  \emph{superactivation} phenomenon, as it implies that both the
  classical and quantum zero-error capacities of these channels can be
  superactivated simultaneously, whilst being a strictly stronger
  property of capacities. Superactivation of the quantum zero-error
  capacity was not previously known.
\end{abstract}


\section{Introduction}
The zero-error capacity, introduced by Shannon in 1956, characterises the
optimal achievable communication rate of a noisy channel when information
must be transmitted with zero probability of
error~\cite{Shannon_zero-error}. This is in contrast with the more
traditional capacity, which only demands error probabilities vanishing in
the limit of many channel uses. The question of zero-error capacity (and
more generally zero-error information theory~\cite{zero-error_review})
has a much more combinatorial flavor than the usual case, and has played
an important role in the development of graph theory. Combinatorial
optimisation problems are often intractable so, perhaps unsurprisingly,
the zero-error capacity is unknown even for many very simple channels.

Quantum information theory seeks to extend information theory to include
information sources and communication systems where quantum effects are
important. Because all physical systems are fundamentally quantum, this
can be seen as an attempt to more accurately model physical information
processing systems. Furthermore, expanding our notion of information to
include quantum messages leads to new insights and applications, such as
quantum cryptography and quantum computing. Because quantum
systems are notoriously delicate, error correction is extremely
important, and the capacities of a noisy quantum channel for transmitting
various types of information noiselessly play a central role in the
theory. In the context of zero-error quantum information theory, first
studied in Ref.~\cite{MA05}, the central capacities are the zero-error
classical and zero-error quantum capacities.

A rather surprising effect has recently been discovered in the theory of
quantum communication. Classically, there is a simple criterion for
deciding whether a channel has non-zero capacity---any channel with some
correlation between input and output has some positive capacity---and
this criterion carries over to the usual classical capacity of quantum
channels. However, when sending quantum information, the situation is
very different. There are some quantum channels that are sufficiently
noisy to have zero capacity for quantum communications, yet can still
create correlations. In Ref.~\cite{graeme+jon} it was shown that there
are pairs of channels with very different noise characteristics, but both
with zero quantum capacity, that, when used together, have a large joint
quantum capacity. This \emph{superactivation} is completely different
from what happens in the classical case, and depends crucially on
choosing entangled signal states for the joint channel.

Superactivation of classical channel capacities is easily seen to be
impossible, both for the usual capacity and the zero-error capacity. If
two classical channels have no correlation between input and output, so
that their usual classical capacity vanishes, this will also hold for the
joint channel. Similarly, if two classical channels each have the
property that all pairs of inputs can lead to ambiguous outputs, so that
the zero-error capacity vanishes, then the joint channel necessarily has
this property too. The argument for the usual classical capacity carries
over directly to the case of quantum channels; superactivation of the
classical capacity of a quantum channel remains impossible.

However, in Ref.~\cite{CCH09} it was shown that the \emph{zero-error}
classical capacity of a quantum channel actually \emph{can} be
superactivated (see also Ref.~\cite{Duan}, which found superactivation of
the non-asymptotic one-shot zero-error classical capacity, and a weaker
form of activation in the asymptotic setting). In this paper, we
significantly strengthen the results and techniques of Ref.~\cite{CCH09}.
There, techniques from algebraic geometry were combined with
probabilistic arguments to show that there are pairs of channels, each
with vanishing zero-error classical capacity, that have positive joint
zero-error classical capacity when used together. Here, we find that
there exist pairs of channels which each have vanishing zero-error
classical capacity, as before, but when the two channels are used
together they can even transmit must more delicate \emph{quantum}
information with zero-error (indeed, only a single use of the joint
channel is required). This is a particularly extreme form of
superactivation, indeed it is the strongest possible form, and has not
been seen previously for other capacities. It implies simultaneous
superactivation of both the classical (already known from
Ref.~\cite{CCH09}) and quantum (previously unknown) zero-error capacities
of quantum channels, whilst being strictly stronger than either of these.

The rest of the paper is organised as follows. In the next section we
review some basic facts about quantum mechanics and algebraic geometry.
\Cref{sec:sufficient} establishes sufficient conditions for this extreme
form of superactivation, whilst \cref{sec:existence} shows that there
exist channels which satisfy these conditions. Finally,
\cref{sec:conclusions} discusses the implications of our findings.

\section{Preliminaries}\label{sec:preliminaries}

\subsection{Quantum Mechanics}
A minimum uncertainty state of a $d$-level quantum system is a pure
state, represented by a $d$-dimensional complex unit vector $\ket{\psi}
\in \CC^d$. More generally, the state of a $d$-level system is given by a
density matrix, $\rho \in \mathcal{B}(\CC^d)$, where,
$\mathcal{B}(\CC^d)$ denotes the set of bounded linear operators on
$\CC^d$. Such a density matrix is Hermitian ($\rho = \rho^\dagger$) and
has unit trace, $\tr\rho = 1$. As a result, any such $\rho$ admits a
spectral decomposition $\rho = \sum_i p_i \proj{\psi_i}$ with orthogonal
$\ket{\psi_i}$, which can be interpreted as describing a system that is
in state $\ket{\psi_i}$ with probability $p_i$. Whilst we will not need
to consider measurement processes below, we will need to know when there
is some measurement to perfectly distinguish two states. This is possible
exactly when the states are orthogonal, i.e.\ for pure states when
$\braket{\psi}{\varphi} = 0$, or for mixed states when $\tr\rho\,\sigma =
0$.

It is sometimes useful to consider (unnormalised) pure states
$\ket[AB]{\psi}$ in a bipartite space $\CC^{d_A}\ox\CC^{d_B}$ as matrices
$M = \mat(\ket[AB]{\psi})$ in the isomorphic space of $d_A\times d_B$
matrices $\mathcal{M}_{d_A,d_B}$. The isomorphism arises from fixing some
product basis $\ket[A]{i}\ket[B]{j}$ for $\CC^{d_A}\ox\CC^{d_B}$, and
expanding $\ket[AB]{\psi} = \sum_{i,j}M_{ij}\ket[A]{i}\ket[B]{j}$ in this
basis. A bipartite subspace $S\subseteq \CC^{d_A}\ox\CC^{d_B}$ is
isomorphic in this way to a matrix subspace which we denote $\mat(S)$.

We define the ``flip'' operation on a bipartite state as the operation
that swaps the two systems and takes the complex conjugate:
\begin{equation}
  \flip(\ket[AB]{\psi}) = \swap(\ket[AB]{\bar{\psi}}).
\end{equation}
In terms of the matrix representation $M = \mat(\ket[AB]{\psi})$, the
flip operation is just Hermitian conjugation: $\mat(\flip\ket[AB]{\psi})
= M^\dg$. The flip operation can be extended to operators and subspaces
in the obvious way.

The most general physical operation in quantum mechanics is a
completely-positive trace preserving (CPT) map from
$\mathcal{B}(\CC^{d_{\mathrm{in}}})$ to
$\mathcal{B}(\CC^{d_{\mathrm{out}}})$, where $d_{\mathrm{in}}$ and
$d_{\mathrm{out}}$ are the input and output dimensions of the map. We
will refer to such operations as \emph{quantum channels} throughout, as
they are directly analogous to channels in classical information theory.
A quantum channel that maps a space $\HS_A$ to $\HS_B$ can always be
thought of as an isometry followed by a partial trace. In other words,
for any channel $\cE$ we have $\cE(\rho) = \mathrm{Tr}_{E}U\rho
U^\dagger$, where $U: \HS_A \rightarrow \HS_B\ox\HS_E$ is an isometry
satisfying $U^\dagger U = I_{A}$. Equivalently, the action of a channel
can be expressed in terms of Kraus operators: ${\cal N}(\rho) = \sum_k
A_k \rho A_k^\dagger$, where $\sum_k A^\dagger_k A_k = I_A$. A third
representation of quantum channels (indeed, it extends to any linear
map), which plays an important role in Ref.~\cite{CCH09}, is the
\keyword{Choi-Jamio\l{}kowski matrix}, defined to be the result of
applying the channel to one half of an unnormalised maximally entangled
state. In other words, the Choi-Jamio\l{}kowski matrix of a channel $\cE$
is given by $\sigma = (\mathcal{I}\otimes\cE)(\omega)$ where
$\ket{\omega}=\sum_{i=1}^{d_A}\ket{i}\ket{i}$, and $\omega =
\proj{\omega}$. The action of the channel can be recovered from the
Choi-Jamio\l{}kowski matrix via $\cE(\rho) =
\tr_A[\sigma_{AB}\cdot\rho_A^T\otimes\id_B]$ (where $\rho_A^T$ denotes
the transpose of the density matrix $\rho_A$).

We will also need the adjoint $\cE^*$ of a channel $\cE$, which is simply
the dual with respect to the Hilbert-Schmidt inner product, i.e.\ the
unique map defined by:
\begin{equation}
  \tr[X^\dagger\,\cE(Y)] = \tr[\,\cE^*(X)^\dagger\,Y].
\end{equation}
In terms of Kraus operators $A_k$, the adjoint $\cE^*$ of $\cE$ is the
map whose Kraus operators are the Hermitian conjugates $A_k^\dg$. (Note
that $\cE^*$ is CP, but not necessarily trace-preserving.)

\subsection{Algebraic Geometry}
In order to prove our results, we need some basic notions from algebraic
geometry (see e.g.\ Ref.~\cite{Harris}). A key concept is that of a
\keyword{Zariski-closed} set, and the resulting \keyword{Zariski
  topology}. We will only ever work over base fields $\CC$ or $\RR$, so
for our purposes Zariski-closed sets are sets defined by a collection of
polynomials, i.e.\ they are the solution sets of simultaneous polynomial
equations. We will use the terms \keyword{Zariski-closed set} and
\keyword{algebraic set} interchangeably.

The Zariski topology is the topology whose \emph{closed} sets are the
Zariski-closed sets. It is the standard topology in algebraic geometry,
but it serves more as a convenient terminology than providing any useful
geometric information. The main use we will make of it is the fact that
intersections of Zariski-closed sets are themselves Zariski-closed.
Indeed, the only Zariski-closed set that has non-zero measure (in the
usual sense on $\CC^d$ or $\RR^d$) is the entire space. This ``Zariski
dichotomy''---that a Zariski-closed set is either zero-measure or the
entire space---lies at the heart of our proofs.

We will also frequently refer to the \keyword{Grassmannian} $\gr_d(V)$ of
a vector space $V$, the set of all $d$-dimensional subspaces of $V$.
There is a standard way of embedding the Grassmannian in projective
space, called the \keyword{Pl\"ucker embedding} and conventionally
denoted $\iota$. If a $d$-dimensional subspace in the Grassmannian is
spanned by some basis $\{\ket{\psi_i}\}$, then $\iota(S)$ is defined to
be $\wedge_{i=1}^d\ket{\psi_i}$, with $\wedge$ denoting the
anti-symmetric product. This is uniquely defined, since picking some
other basis replaces $\ket{\psi_i}$ by $\sum_{j=1}^d A_{i,j}
\ket{\psi_j}$ for some invertible matrix $A$, which in turn replaces
$\iota(S)$ by $\det(A)\iota(S)$. In projective space, rescaling by the
scalar $\det(A)$ makes no difference.

Via the Pl\"ucker embedding, points in the Grassmannian are naturally
parametrised by the coordinates of points in projective space, called
the \keyword{Pl\"ucker coordinates}. (Note that not all points in the
ambient projective space correspond to points in the Grassmannian; the
Pl\"ucker coordinates of points within the Grassmannian must satisfy
quadratic constraints called the \keyword{Pl\"ucker relations}.) Thus the
Pl\"ucker coordinates $P_{\alpha_d}$ of $S$ are defined by
$\sum_{\alpha_d}P_{\alpha_d}(\wedge_{j \in \alpha_d}\ket{j}) =
\wedge_{i=1}^d\ket{\psi_i}$, where $\alpha_d$ are size $d$ subsets of
$\{1\dots n\}$, with $\ket{1},\dots,\ket{n}$ a basis of $V$.

\section{Sufficient Conditions for Superactivation}
\label{sec:sufficient}
We start by reducing the problem of proving existence of our extreme form
of superactivation to a question about the existence of subspaces
satisfying certain conditions. The arguments are very similar to those
leading to Theorem~13 of Ref.~\cite{CCH09}, but the stronger requirement
that the joint channel have positive \emph{quantum} zero-error capacity
adds an additional constrain on the subspaces. To derive this new
constraint, we need the following lemma, which gives us a sufficient
condition for a channel to have positive zero-error quantum capacity.

\begin{lemma}\label{lem:quantum_zero-error}
  Let $\cE:\HS_A \rightarrow \HS_B$ be a channel, $\ket{0}$ and $\ket{1}$
  be states on $\HS_A$, and $\ket{{\pm}} = 1/\sqrt{2}(\ket{0} \pm
  \ket{1})$. Then, if
  \begin{equation}
    \tr\left[ \cE(\proj{0})\,\cE(\proj{1}) \right] = 0
  \end{equation}
  and
  \begin{equation}
    \tr\left[ \cE(\proj{+})\,\cE(\proj{-}) \right] = 0,
  \end{equation}
  we have $Q_0(\cE) \geq 1$.
\end{lemma}
\begin{IEEEproof}
  To see this, suppose $A_k$ are the Kraus operators of $\cE$ and
  $\varphi = \frac{1}{2}(\proj{0}+\proj{1})$, and let
  \begin{equation}
    \cR_{\varphi}(\rho)
    = \sum_{k}\sqrt{\varphi}A^\dg_k \cE(\varphi)^{-1/2}\rho\,
              \cE(\varphi)^{-1/2} A_k \sqrt{\varphi} + \Pi\rho \Pi,
  \end{equation}
  where $\cE(\varphi)^{-1/2}$ is the square-root of the Moore-Penrose
  pseudo-inverse of $\cE(\varphi)$ (i.e.\ its inverse when restricted to
  its support), and $\Pi$ is the projector onto the kernel of
  $\cE(\varphi)$ (which vanishes if $\cE(\varphi)$ is invertible). This
  corresponds to the reversal operation of Ref.~\cite{Barnum} when
  $\cE(\varphi)$ is full rank. It is completely-positive and trace
  preserving by design, and $\cM := \cR_\varphi \circ \cE$ is the
  identity on $\vspan(\ket{0},\ket{1})$.

  To see this, first note that, by assumption,
  \begin{subequations}
  \begin{align}
    0 &= \tr\left[\cE(\proj{0})\cE(\proj{1})\right]\\
      &= \sum_{j,k}\tr\left[A_j\proj{0}A_j^\dg A_k \proj{1}A_k^\dg\right]\\
      &= \sum_{j,k}\Abs{\bra{0}A_j^\dg A_k \ket{1}}^2,
  \end{align}
  \end{subequations}
  so that $\bra{0}A_j^\dg A_k \ket{1} = 0$ for all $j,k$, and similarly
  for $\bra{+}A_j^\dg A_k\ket{-}$. Now consider
  \begin{subequations}\label{Eqs:As}
  \begin{align}
    &\tr\left[
      \sqrt{\varphi}A_k^\dg \cE(\varphi)^{-1/2} \cE(\proj{0})
      \cE(\varphi)^{-1/2} A_k \sqrt{\varphi} \proj{1}
    \right]\\
    &=\frac{1}{2}\tr\left[
        A_k^\dg \cE(\varphi)^{-1/2} \cE(\proj{0})
        \cE(\varphi)^{-1/2} A_k \proj{1}
      \right]\\
    &=\frac{1}{2}\tr\left[
        \cE(\varphi)^{-1/2} \cE(\proj{0})
        \cE(\varphi)^{-1/2} \cE(\proj{1})
      \right].
  \end{align}
  \end{subequations}
  Since $\cE(\proj{0})$ and $\cE(\proj{1})$ are orthogonal, we have
  \begin{equation}
    \cE(\varphi)^{-1/2}
    = \sqrt{2}\cE(\proj{0})^{-1/2} + \sqrt{2}\cE(\proj{1})^{-1/2},
  \end{equation}
  which immediately implies with Eq.~(\ref{Eqs:As}) and the fact that $\cE(\proj{0})$ and $\cE(\proj{1})$ are in
the support of $\cE(\varphi)$ that
  \begin{subequations}
  \begin{align}
    \cM(\proj{0}) &= \proj{0}\\
    \cM(\proj{1}) &= \proj{1}.
  \end{align}
  \end{subequations}
  Similarly, we also have
  \begin{subequations}
  \begin{align}
    \cM(\proj{+}) &= \proj{+}\\
    \cM(\proj{-}) &= \proj{-}.
  \end{align}
  \end{subequations}
  Now all we have to do is show that any CPT map $\cM$ satisfying the
  above four equations must be the identity. We can easily use these four
  equations to show that
  \begin{subequations}
  \begin{align}\label{eq:ids}
    \cM(\id) &= \id,\\
    \cM(X) &= X,\\
    \cM(Z) &= Z,
  \end{align}
  \end{subequations}
  where $X = \left (\begin{smallmatrix}0 & 1\\ 1 &
      0\end{smallmatrix}\right)$ and $Z = \left (\begin{smallmatrix}1 &
      \phantom{-}0\\ 0 & -1
    \end{smallmatrix}\right)$.

Since $\cM$ is a unital qubit channel \cite{KR99}, it is a mixture of conjugations of Pauli
matrices of the form
\begin{equation}
\cM(\rho) = (1-p_X-p_Y-p_Z)\rho + p_X X\rho X+p_Y Y\rho Y+p_Z Z\rho Z,
\end{equation}
where $Y = \left(\begin{smallmatrix}0 & -i \\
    i & \phantom{-}0 \end{smallmatrix}\right)$. This form, together with
\cref{eq:ids}, implies that $p_X=p_Y=p_Z = 0$, so that $\cM(\rho) =
\rho$.
\end{IEEEproof}

We are now in a position to reduce our superactivation problem to a
question about subspaces. The approach is the closely related to that in
Ref.~\cite{CCH09}, which in turn builds on the techniques of
Ref.~\cite{rank_additivity}. We start by recapping the conditions
required for superactivation of the \emph{classical} zero-error capacity
from Ref.~\cite{CCH09}, which is necessary (but not sufficient) for our
result. We then show how to strengthen this to achieve the extreme form
of superactivation claimed here.

Recall that two quantum states $\rho,\sigma$ are perfectly
distinguishable if an only if they are orthogonal ($\tr[\rho\sigma]=0$).
Thus, the \emph{classical} zero-error capacity of a channel $\cE$ is 0
iff no pair of inputs gives orthogonal outputs:
\begin{equation}\label{eq:individual}
  \forall\psi,\varphi: \quad
  0 \neq \tr[\cE(\varphi)\cE(\psi)]
    = \tr[\psi\cdot\cE^*\circ\cE(\varphi)],
\end{equation}
where we have simply pulled the channel across the inner product in the
final equality, giving the composition of the adjoint $\cE^*$ and the
channel.
% Conversely, the classical zero-error capacity is non-zero iff
% \begin{equation}
%   \exists\psi,\varphi \quad
%   \tr[\psi\cdot\cE^*\circ\cE(\varphi)] = 0.
% \end{equation}
Rewriting these expressions by expressing the action of the composite map
$\cN=\cE^*\circ\cE$ in terms of its Choi-Jamio\l{}kowski state $\sigma$,
this is equivalent to:
\begin{equation}
  \forall\psi,\varphi: \quad
  \tr\left[\sigma\cdot\varphi_A^T\otimes\psi_{A'}\right] \neq 0.
\end{equation}
But this simply expresses the condition that $\sigma$ should not be
orthogonal to any product state. Therefore, for the channel $\cE$ to have
no classical zero-error capacity, the support $S$ of $\sigma$ must
contain no product states in its orthogonal complement $S^\perp$.

Thus in order to superactivate the \emph{one-shot, classical} zero-error
capacity, we need two subspaces $S_1,S_2$ (corresponding to two channels
$\cE_1,\cE_2$ as described above), each of which has no product states in
its orthogonal complement, such that the joint channel $\cE_1\ox\cE_2$
\emph{does} have positive classical zero-error capacity. To achieve
superactivation even in the asymptotic setting, we must strengthen the
condition on the individual subspaces to ensure that even arbitrarily
many copies of the individual channels have no capacity. Since the
Choi-Jamio\l{}kowski matrix of $k$ copies $\cN^{\ox k}$ of a map is the
tensor power $\sigma^{\ox k}$ of the single-copy Choi-Jamio\l{}kowski
matrix, this is equivalent to requiring that no \emph{tensor power}
$S_{1,2}^{\ox k}$ of either subspace has a product state in its
orthogonal complement.

As well as the individual channels having no capacity, we also want the
joint channel $\cE_1\ox\cE_2$  to have positive zero-error capacity,
i.e.\ we require the converse of \cref{eq:individual} to hold for the
joint channel:
\begin{equation}\label{eq:joint}
  \exists \psi,\varphi: \tr[\psi\cdot
  (\cE_1^*\circ\cE_1)\otimes(\cE_2^*\circ\cE_2)(\varphi)] = 0.
\end{equation}
Let us choose the (unnormalised) inputs $\ket{\psi},\ket{\varphi}$ to the
joint channel to be the maximally entangled states $\ket{\omega}$ and
$\id\ox X\ket{\omega}$, where $X$ is now the generalisation of the Pauli
$X$ matrix to arbitrary dimension, i.e.\ the matrix with ones down its
anti-diagonal. Expressing \cref{eq:joint} in terms of the
Choi-Jamio\l{}kowski matrix $\sigma_1\ox\sigma_2$ of the joint channel,
the condition of \cref{eq:joint} simplifies to:
\begin{equation}
  \tr\left[
    \sigma_1^T \cdot (\id\ox X)\,\sigma_2\,(\id\ox X^\dg)
  \right] = 0.
\end{equation}
This simply expresses the condition that
$(X\ox\id)\,\sigma_2\,(X^\dg\ox\id)$ and $\sigma_1^T$ should have
orthogonal supports, i.e.\ $(X\ox\id) S_2 \perp S_1^T$. Since we also
want the individual subspaces to have no product states in their
orthogonal complements, it makes sense to choose the two subspaces to be
as big as possible (so that their orthogonal complements are as small as
possible), subject to this condition. We therefore choose $S_2$ to be the
orthogonal complement (up to the local unitary rotation and
transposition) of $S_1$:
\begin{equation}\label{eq:orthogonal_complement}
  S_2^T = \id\otimes X\cdot S_1^\perp.
\end{equation}
This allows us to express all the requirements for classical zero-error
superactivation in terms of conditions on a \emph{single} subspace
$S:=S_1$. These conditions are summarised in the following theorem (which
is Theorem~13 from Ref.~\cite{CCH09}):
\begin{theorem}\label{thm:classical_S_conditions}
  If there exists a subspace $S$ and unitaries $U,V$ satisfying
  \begin{subequations}
  \label[equations]{eq:S_strongly_unextendible_conditions}
  \begin{gather}
    \forall k,\nexists\ket{\psi},\ket{\varphi}\in\HS_A^{\otimes k}:
      \ket{\psi}\otimes\ket{\varphi} \in (S^{\otimes k})^\perp,\\
    \forall k,\nexists\ket{\psi},\ket{\varphi}\in\HS_A^{\otimes k}:
      \ket{\psi}\otimes \ket{\varphi}
      \in \bigl((S^\perp)^{\otimes k}\bigr)^\perp,\\
    \flip(S) = S\,,\label{eq:symmetry1}\\
    \flip(\id\otimes X\cdot S) = \id\otimes X\cdot S,\\
    \exists \{M_i \geq 0\}: \mat(S) = \vspan\{M_i\},\\
    \exists \{M_j \geq 0\}: \mat(\id\otimes X\cdot S^\perp) =
    \vspan\{M_j\},
      \label{eq:symmetry4}
  \end{gather}
  \end{subequations}
  then there exist channels $\cE_{1,2}$ with $C_0(\cE_1) = C_0(\cE_2) =
  0$ but $C_0(\cE_1\ox\cE_2) \geq 1$.
\end{theorem}
The final four conditions in \crefrange{eq:symmetry1}{eq:symmetry4}
express the requirement that the subspace must come from the support of a
Choi-Jamio\l{}kowski matrix of a composite map with the very particular
form $\cE^*\circ\cE$, which imposes additional symmetries on the
subspace. (Fuller details of the proof can be found in
Ref.~\cite[Theorem~13]{CCH09}.)

\Cref{thm:classical_S_conditions} gives sufficient conditions for
superactivation of the \emph{classical} zero-error capacity. But we want
something significantly stronger; we not only want the joint channel to
have positive classical zero-error capacity, we want it even to have
positive \emph{quantum} zero-error capacity. For this, we must strengthen
\Cref{thm:classical_S_conditions} using \cref{lem:quantum_zero-error}:
\begin{theorem}\label{thm:S_strongly_unextendible_conditions}
  Suppose there is a subspace $S$ of a bipartite Hilbert space
  \mbox{$\HS_A\ox\HS_A$} such that
  \begin{subequations}
  \begin{gather}
    \forall k,\;\nexists\ket{\psi},\ket{\varphi}\in\HS_A^{\ox k}:
      \ket{\psi}\ox\ket{\varphi} \in (S^{\ox k})^\perp,
      \label{eq:S_strongly_unextendible}\\
    \forall k,\;\nexists\ket{\psi},\ket{\varphi}\in\HS_A^{\ox k}:
      \ket{\psi}\ox \ket{\varphi}
      \in \bigl((S^\perp)^{\ox k}\bigr)^\perp,
      \label{eq:Sperp_strongly_unextendible}\\
    \flip(S) = S\,,\label{eq:S_conjugate_symmetry}\\
    \flip(\id\ox X\cdot S) = \id\ox X\cdot S,
      \label{eq:Sperp_conjugate_symmetry}\\
    \exists \{M_i \geq 0\}: \mat(S) = \vspan\{M_i\},
      \label{eq:S_positivity}\\
    \exists \{M_j \geq 0\}: \mat(\id\ox X\cdot S^\perp) = \vspan\{M_j\},
      \label{eq:Sperp_positivity}\\
    S \perp (\id+X)\ox (\id-X) S^\perp.\label{eq:S_orthogonality}
  \end{gather}
  \end{subequations}
  Then there exist channels $\cE_{1,2}$ with $C_0(\cE_1) = C_0(\cE_2) =
  0$ but $Q_0(\cE_1\ox\cE_2)\geq 1$.
\end{theorem}
\begin{IEEEproof}
  \Crefrange{eq:S_strongly_unextendible}{eq:Sperp_positivity} are
  identical to the conditions in \cref{thm:classical_S_conditions}, and
  already give sufficient conditions for the individual channels to have
  no zero-error capacity, $C_0(\cE_1) = C_0(\cE_2) = 0$, and the joint
  channel to have positive \emph{classical} zero-error capacity,
  $C_0(\cE_1\ox\cE_2) \geq 1$. Only \cref{eq:S_orthogonality} is new. We
  must show that this additional condition is sufficient to ensure the
  joint channel has positive \emph{quantum} zero-error capacity,
  $Q_0(\cE_1\ox\cE_2)\geq 1$.

  Recall from \cref{thm:classical_S_conditions} and Ref.~\cite{CCH09}
  that $S$ will be the support of $\sigma_1 = (\mathcal{I} \ox
  \cE_1^*\circ \cE_1)(\omega)$ with $\ket{\omega} =
  \sum_{i}\ket{i}\ket{i}$ and $S_2 = (\id\ox X)S^\perp$ the support of
  $\sigma_2^T$ defined similarly. The two signal states for $\cE_1 \ox
  \cE_2$ in \cref{thm:classical_S_conditions} are $\ket{\varphi_0} =
  \ket{\omega}$ and $\ket{\varphi_1} = (\id\ox X)\ket{\omega}$. From
  \cref{lem:quantum_zero-error}, what we have to do now is show that,
  letting $\ket{\varphi_\pm} =
  (\ket{\varphi_0}\pm\ket{\varphi_1})/\sqrt{2}$, we have
  \begin{equation}\label{eq:PlusMinusZero}
    \tr\left[(\cE_1 \ox \cE_2) (\proj{\varphi_+})
             (\cE_1 \ox \cE_2)(\proj{\varphi_-})\right] = 0.
  \end{equation}
  Now,
  \begin{subequations}
  \begin{align}
    &\tr\bigl[(\cE_1 \ox \cE_2) (\proj{\varphi_+})
              (\cE_1 \ox \cE_2)(\proj{\varphi_-})\bigr]\\[0.5em]
    &=\tr\bigl[(\cE_1^*\circ\cE_1 \ox \cE_2^*\circ\cE_2)
               (\proj{\varphi_+})\proj{\varphi_-}\bigr]\\[0.5em]
    &=\tr\left[\sigma_1^{A_1'A_1} \ox \sigma_2^{A_2'A_2}
               \proj[A_1'A_2']{\varphi_+}^T \ox
               \proj[A_1A_2]{\varphi_-}\right]\\[0.5em]
    &\begin{aligned}
      =\tr\biggl[\sigma_1^{A_1'A_1} \ox \sigma_2^{A_2'A_2}\;
                 &\Bigl(
                   P_+\ox\id \proj[A_1'A_2']{\omega} P_+\ox\id
                 \Bigr)\;\\[-0.7em]
                 &\mspace{20mu}\ox\Bigl(
                   P_-\ox\id \proj[A_1A_2]{\omega} P_-\ox\id
                 \Bigr)
          \biggr]
      \end{aligned}\\
    &\begin{aligned}
      =\tr\biggl[(P_+\ox P_-) \sigma_1^{A_1'A_1} &(P_+\ox P_-) \ox
               \sigma_2^{A_2'A_2}\cdot\\
               &\mspace{63mu}\proj[A_1'A_2']{\omega} \ox
               \proj[A_1A_2]{\omega}\biggr]
    \end{aligned}\\
    &=\tr\left[\left[(P_+\ox P_-) \sigma_2 (P_+\ox P_-)\right]^T
               \sigma_1\right]
  \end{align}
  \end{subequations}
  where $P_\pm = (\id\pm X)/2$ are projectors and we have used the fact
  that $\tr[\proj{\omega} M \ox N]= \tr[N^T M]$. As a result, the
  requirement \cref{eq:PlusMinusZero} is met by choosing $S \perp (P_+\ox
  P_-)S_2$. This is equivalent to $S \perp (\id+X)\ox (\id-X)S^{\perp}$,
  since $P_- X = -P_-$ and we chose $S_2 = (\id\ox X) S^\perp$.
\end{IEEEproof}

\section{Existence of Superactivation}
\label{sec:existence}
Given \cref{thm:S_strongly_unextendible_conditions}, all we need to do in
order to show the extreme superactivation phenomenon is to prove that
there do exist subspaces satisfying the conditions of the theorem. We use
a combination of algebraic-geometry and probabilistic arguments to
establish this result.

In what follows, we will need to consider a number of sets of subspaces.
Recall the definition of extendibility from
Refs.~\cite{CCH09,Jianxin_strong}:
\begin{definition}
  A subspace $S\subseteq\HS_A\otimes\HS_B$ is $k$-unextendible if
  $(S^{\otimes k})^\perp$ contains no product state in $\HS_{A^{\otimes
      k}}\otimes\HS_{B^{\otimes k}}$. A subspace is \keyword{strongly
    unextendible} if it is $k$-unextendible for all $k\geq 1$.
  Conversely, a subspace is \keyword{$k$-extendible} if it is not
  $k$-unextendible, and \keyword{extendible} if it is not strongly
  unextendible.
\end{definition}
Following Ref.~\cite{CCH09}, we denote the sets of $d$-dimensional
$k$-extendible, extendible, and unextendible subspaces, respectively, by
\begin{align}
  E_d^k(\HS_A,\HS_B)
    &=\{S\in\gr_d(\HS_A\ox\HS_B)
        | S \text{ is $k$-extendible}\},\\
  E_d(\HS_A,\HS_B)
    &=\{S\in\gr_d(\HS_A\ox\HS_B)
        | S \text{ is extendible}\},\\
  U_d(\HS_A,\HS_B)
    &=\{S\in\gr_d(\HS_A\ox\HS_B)
        | S \text{ is unextendible}\},
\end{align}

Note that the set $\bigcup_k E_d^k$ is the set of subspaces that
\emph{do} contain product states in their orthogonal complements, so it
is precisely the set of subspaces that we want to avoid in order to
satify the condition in \cref{eq:S_strongly_unextendible}. At the heart
of our proof is the following Lemma, which shows that the set $E_d^k$
algebraic:
\begin{lemma}\label{lem:Edk-closed}
  $E_d^k(\HS_A,\HS_B)$ is Zariski-closed in
  $\gr_d(\HS_A\otimes\HS_B) =
  \gr_d(\CC^{d_A}\otimes\CC^{d_B})$.
\end{lemma}
This is proven in Lemma~15 of Ref.~\cite{CCH09} using standard algebraic
geometry arguments, based onthe fact that there is a simple algebraic
characterisation of product states $\ket[AB]{\psi}$ as those states for
which $\mat(\ket[AB]{\psi})$ is rank~1.

We will also refer to the set
\begin{multline}\label{eq:Fd-real}
  F_d(\RR,d_A)
  = \{S \in \gr_{2d}(\RR^2\ox\RR^{d_A}\ox\RR^{d_A}) \,|\\
      S=iS, \flip(S) = S,\; \flip(\id\ox X\cdot S) = \id\ox X \cdot S\}
\end{multline}
of subspaces satisfying the symmetry constraints of
\cref{eq:S_conjugate_symmetry,eq:Sperp_conjugate_symmetry}. Note that we
are considering $F_d$ as a subset of the \emph{real} Grassmannian, in
which context $i=\left(\begin{smallmatrix} 0 & -1 \\ 1 & \phantom{-}0
\end{smallmatrix}\right)$.

\begin{lemma}\label{lem:Fd_Zariski_closed}
  $F_d(\RR,d_A)$ is Zariski-closed in
  $\gr_{2d}(\RR^2\otimes\RR^{d_A}\otimes\RR^{d_A})$.
\end{lemma}
This is proven in Lemma~17 of Ref.~\cite{CCH09}, writing out the
constraints on $S$ from \cref{eq:Fd-real} explicitly in terms of the
Pl\"ucker coordinates, and verifying that the constraints are
polynomials.

In order to extend the arguments of Ref.~\cite{CCH09} to our case, we
will need to consider an additional set: the set of subspaces satisfying
the orthogonality constraint of \cref{eq:S_orthogonality}:
\begin{multline}
  C_d(\CC,d_A)
  = \{S \in \gr_d(\CC^{d_A}\ox\CC^{d_A})\\
      | S \perp (\id+X)\ox(\id-X)S^\perp\},
\end{multline}
and also the isomorphic set of real vector spaces:
\begin{equation}\label{eq:Cd-real}
  \begin{split}
    C_d(\RR,d_A)
    &= \{S \in \gr_{2d}(\RR^2\ox\RR^{d_A}\ox\RR^{d_A})\\
        &\mspace{60mu} |S=iS,\; S \perp (\id+X)\ox(\id-X)S^\perp\}.
  \end{split}
\end{equation}

\vspace{1em}\noindent
The first step is to show that this set is algebraic (cf.\ Lemma~17 of
Ref.~\cite{CCH09}).
\begin{lemma}
  $C_d(\CC,d_A)$ is Zariski-closed in $\gr_d(\CC^{d_A}\ox\CC^{d_A})$.
\end{lemma}
\begin{IEEEproof}
  First, we let $W = \wedge_{i=1}^{d}\ket{\psi_i}$ for some basis
  $\{\ket{\psi_i}\}$ of $S$. We have $\ket{\psi} \in S$ exactly when
  $\ket{\psi}\wedge W = 0$ and we want to use this to construct a basis
  for $S^\perp$. If $P_{\alpha_d}$ are the Pl\"ucker coordinates of $S$,
  and supposing $\ket{\psi} = \sum_{i=1}^n v_i\ket{i}$, then $\ket{\psi}$
  is in $S$ exactly when
  \begin{subequations}
  \begin{align}
    \ket{\psi}\wedge W
    &=\sum_{\alpha_d}\sum_i v_i
        P_{\alpha_d} \ket{i} \wedge (\wedge_{j \in \alpha_d}\ket{j})\\
    &=\sum_{i,\beta_{d+1}}v_i N_{i,\beta_{d+1}} \wedge_{k \in \beta_{d+1}}\ket{k}
     = 0,
  \end{align}
  \end{subequations}
  so that we have an $N$ such that $\ket{\psi} \in S$ iff $\bra{\psi}N =
  0$. Now, the support of $N N ^\dg$ is $S^\perp$ and its eigenvalues are
  positive. Most importantly, we can think of $NN^\dg$ as a matrix with
  entries that are quadratic polynomials in $P_{\alpha_d}$. Thus, we are
  interested in ensuring that
  \begin{equation}
    N\cdot P_+ \ox P_- \ket{\psi} = 0
  \end{equation}
  for all $\ket{\psi} \in S$, which is equivalent to showing that
  \begin{equation}
    N^{\ox n}(P_+ \ox P_-)^{\ox n}\iota(S) = 0.
  \end{equation}
  This is a linear constraint on $\iota(S)$, so $\{\iota(S) : N^{\ox
    n}(P_+ \ox P_-)^{\ox n}\iota(S) = 0\}$ is Zariski-closed. Since
  $\iota$ is a proper morphism (cf.\ Lemma~17 of Ref.~\cite{CCH09}), $C_d
  = \{S : \forall\ket{\psi}\in S, N\cdot P_+ \ox P_- \ket{\psi} = 0\}$
  must also be Zariski-closed.
\end{IEEEproof}

Any Zariski-closed set in a complex vector space is also Zariski-closed
in the isomorphic real vector space. Furthermore, the intersection of two
Zariski-closed sets is again Zariski-closed, since they form a topology.
This immediately gives:
\begin{corollary}\label{cor:Ed_Fd_Cd_Zariski-closed}
  $E_d^k(\HS_A,\HS_{A'}) \cap F_d(\RR,d_A) \cap C_d(\RR,d_A)$ is
  Zariski-closed in $F_d(\RR,d_A) \cap C_d(\RR,d_A)$.
\end{corollary}

We can now use the ``Zariski dichotomy'' to prove that the set of
strongly unextendible subspaces is full measure in $F_d\cap C_d$. Note
that our results are not particularly sensitive to the choice of measure,
but for definiteness, when we refer to a measure or to a probability
distribution on the Grassmannian, this can always be taken to be the one
induced by the Haar measure over the unitary group. More explicitly, the
action of the unitary group on a Hilbert space induces a natural measure
on quantum states --- the standard choice in quantum information theory.
This in turn induces a measure on subspaces of a given dimension, i.e.\
on the Grassmannian. When refer to a measure on a subset of the
Grassmannian, we mean the restriction of the measure over the whole
Grassmannian to that subset.

We will make use of \keyword{unextendible product bases} in the proofs,
which are defined as follows:
\begin{definition}
  An \keyword{unextendible product basis} (UPB) is a set of product
  states $\{\ket[AB]{\psi_i}\}$ (not necessarily orthogonal) in a
  bipartite space $\HS_A\otimes\HS_B$ such that
  $(\vspan\{\ket{\psi_i}\})^\perp$ contains no product states. The
  \keyword{dimension} of a UPB is the number of product states in the
  set.
\end{definition}
Clearly, a UPB spans a $1$-unextendible subspace. In fact, Lemma~22 of
Ref.~\cite{CCH09}, which we restate here, shows that the span is even
strongly unextendible:
\begin{lemma}\label{lem:UPB_strong_unextendible}
  If $\{\ket[A_1B_1]{\psi_i^1}\}$ and $\{\ket[A_2B_2]{\psi_i^2}\}$ are
  unextendible product bases in $\HS_{A_1}\otimes\HS_{B_1}$ and
  $\HS_{A_2}\otimes\HS_{B_2}$ respectively, then
  $\{\ket{\psi_i^1}\ket{\psi_j^2}\}_{i,j}$ is an unextendible product
  basis in $\HS_{A_1A_2}\otimes\HS_{B_1B_2}$.
\end{lemma}

We are now in a position to prove the following key lemma.
\begin{lemma}\label{lem:Ud_full-measure}
  For $d \geq 12(d_A+d_B-1)$, the set of strongly unextendible subspaces
  $U_d(\HS_A,\HS_{A'}) \cap F_d(\CC,d_A) \cap C_d(\CC,d_A)$ is full
  measure in $F_d(\CC,d_A) \cap C_d(\CC,d_A)$.
\end{lemma}
\begin{IEEEproof}
  Since $E_d^k(\HS_A,\HS_{A'}) \cap F_d(\RR,d_A) \cap C_d(\RR,d_A)$ is
  Zariski-closed by \cref{cor:Ed_Fd_Cd_Zariski-closed},
  $\bigcup_kE_d^k(\HS_A,\HS_{A'}) \cap F_d(\RR,d_A) \cap C_d(\RR,d_A)$ is
  a countable union of Zariski-closed sets, so it is either zero measure
  in $F_d(\RR,d_A) \cap C_d(\RR,d_A)$, or it is the full space.
  Conversely, its complement $U_d(\HS_A,\HS_{A'}) \cap F_d(\RR,d_A) \cap
  C_d(\RR,d_A)$ is either full measure or empty.

  To rule out the possibility that it is empty, we prove that there
  exists a subspace in $U_d \cap F_d \cap C_d$ by constructing one using
  unextendible product bases (UPBs). \Cref{lem:UPB_strong_unextendible}
  shows that the span of a UPB is a strongly unextendible subspace, and
  it is known from Ref.~\cite{Bhat04} that UPBs of dimension $m$ exist in
  $\CC^{d_A}\ox\CC^{d_B}$ for any $m \geq d_A + d_B - 1$. Let $S$ be a
  subspace spanned by such a minimal UPB, and let the set of matrices
  $\{M_i\}$ be a basis for $\mat(S)$. Consider the symmetrised subspace
  $\mat(S')$ spanned by
  \begin{equation}
    \begin{split}
      \Bigl\{
        &M,\; XMX,\; M^\dg,\; XM^\dg X, \\
        &\quad P_+MP_-,\; P_+XMXP_-,\; P_+M^\dg P_-,\; P_+XM^\dg XP_-, \\
        &\qquad P_-MP_+,\; P_-XMXP_+,\; P_-M^\dg P_+,\; P_-XM^\dg XP_+
      \Bigr\}.
    \end{split}
  \end{equation}
  The resulting subspace $S'$ has dimension at most $12(d_A+d_B-1)$, and
  satisfies both the symmetry and orthogonality constraints of
  \cref{eq:S_conjugate_symmetry,eq:Sperp_conjugate_symmetry,eq:S_orthogonality}
  from \cref{thm:S_strongly_unextendible_conditions}. Thus $S' \in
  F_d\cap C_d$. Since $S$ is strongly-unextendible, and $S \subseteq S'$,
  $S'$ is clearly strongly unextendible, which completes the proof.
\end{IEEEproof}

\begin{corollary}\label{cor:symmetric_strongly_unextendible}
  For any $d_A \geq 48$, and for a subspace $S\in\CC^{d_A}\ox\CC^{d_A}$
  of dimension $12(2d_A-1) \leq d \leq d_A^2 - 12(2d_A-1)$ chosen at
  random\footnote{E.g.\ according to the distribution induced by the Haar
    measure; see discussion preceding \cref{lem:Ud_full-measure}.}
  subject to the constraints $\flip(S)=S$, $\flip(\id\ox X\cdot S) =
  \id\ox X\cdot S$ and $S \perp (\id+X)\ox(\id-X)\,S^\perp$, both $S$ and
  $S^\perp$ will almost-surely be strongly unextendible.
\end{corollary}
\begin{IEEEproof}
  \Cref{lem:Ud_full-measure} implies that $S$ chosen in this way will
  almost-surely be strongly unextendible. But $S^\perp$ is then a random
  subspace subject to the same constraints, with dimension $12(2d_A-1)
  \leq d^\perp = d_A^2 - d \leq d_A^2 - 12(2d_A-1)$. Thus
  \cref{lem:Ud_full-measure} implies that $S^\perp$ will also be
  almost-surely strongly unextendible. For there to exist a suitable $d$,
  we require $12(2d_A-1) \leq d_A^2 - 12(2d_A-1)$, or $d_A \geq 48$.
\end{IEEEproof}

\Cref{cor:Ed_Fd_Cd_Zariski-closed} tells us that, although
\cref{eq:S_strongly_unextendible,eq:Sperp_strongly_unextendible} of
\cref{thm:S_strongly_unextendible_conditions} would appear to impose
severe constraints on the subspace $S$, they are in fact benign. Even if
we restrict to subspaces satisfying
\cref{eq:S_conjugate_symmetry,eq:Sperp_conjugate_symmetry,eq:S_orthogonality},
a randomly chosen subspace will satisfy
\cref{eq:S_strongly_unextendible,eq:Sperp_strongly_unextendible} with
probability~1.

It remains to show that such a subspace can also satisfy
\cref{eq:S_positivity,eq:Sperp_positivity}. For this, we require more
information about the structure of the set $F_d\cap C_d$ of subspaces
that simultaneously satisfy
\cref{eq:S_conjugate_symmetry,eq:Sperp_conjugate_symmetry,eq:S_orthogonality}.
\begin{lemma}\label{lem:Fd-Cd_structure}
  If $d_A$ is even, then
  \begin{equation}
    \begin{split}
      &F_d(\RR,d_A)\cap C_d(\RR,d_A)\\
      &\cong\mspace{10mu}
        \bigsqcup_{\mathllap{r=\max[0,d-\frac{d_A^2}{2}]}}^{\min[d,\frac{d_A^2}{2}]}\quad
        \bigsqcup_{k_1=0}^{r}\;\; \bigsqcup_{k_2=0}^{d-r}
        \Bigl(\gr_{k_1}(\RR^{d_A^2/2}) \times\\
      &\mspace{40mu}
        \gr_{r-k_1}(\RR^{d_A^2/2}) \times
        \gr_{k_2}(\RR^{d_A^2/2}) \times \gr_{d-r-k_2}(\RR^{d_A^2/2})\Bigr).
    \end{split}\raisetag{4em}
  \end{equation}
  The $\sqcup$ denotes disjoint union, meaning an element of
  $F_d(\RR,d_A)\cap C_d(\RR,d_A)$ can be uniquely identified by
  specifying non-negative integers $r$, $k_1$ and $k_2$ satisfying
  $d-d_A^2/2 \leq r \leq d$, $k_1 \leq r$ and $k_2 \leq d-r$, along with
  elements of $\gr_{k_1}(\RR^{d_A^2/2})$, $\gr_{r-k_1}(\RR^{d_A^2/2})$,
  $\gr_{k_2}(\RR^{d_A^2/2})$ and $\gr_{d-r-k_2}(\RR^{d_A^2/2})$.
\end{lemma}
\begin{IEEEproof}
  Elements of $F_d(\RR,d_A)\cap C_d(\RR,d_A)$ are $2d$-dimensional real
  subspaces of $\RR^2 \ox \RR^{d_A} \ox \RR^{d_A}$. As such, they can be
  expressed as rank-$2d$ projectors. In terms of these projectors $\Pi$,
  the constraints in \cref{eq:Fd-real} defining $F_d(\RR,d_A)$ become
  $i\,\Pi\,i^T = \Pi$, $\flip\,\Pi\,\flip^T=\Pi$ and $(X \ox X)\Pi (X\ox
  X)= \Pi$ (cf.\ Lemma~28 of Ref.~\cite{CCH09}).

  The additional constraint $S \perp (P_+\ox P_-) S^\perp$ in
  \cref{eq:Cd-real} defining $C_d(\RR,d_A)$ can also be expressed as a
  symmetry of $\Pi$. Note that this constraint is symmetric; if $S$
  satisfies it, then so does $S^\perp$. To see this, express the
  constraint as
  \begin{equation}
    \forall \ket{\psi}\in S^\perp,\ket{\varphi}\in S:
    \BraKet{\varphi}{P_+\otimes P_-}{\psi} = 0
  \end{equation}
  and take the complex conjugate. If $\Pi$ is the projector corresponding
  to a subspace $S$, the constraint is equivalent to
  \begin{equation}\label{eq:Cd_constraint}
    \Pi (P_+\ox P_-) \Pi = (P_+\ox P_-)\Pi,
  \end{equation}
  and we know the same holds for $S^\perp$:
  \begin{equation}
    (\id-\Pi) P (\id-\Pi) = P (\id-\Pi),
  \end{equation}
  % From the latter, we obtain
  % \begin{equation}
  %   (P_+\ox P_-) - (P_+\ox P_-)\Pi
  %   = (P_+\ox P_-) (\id-\Pi)
  %   = (\id-\Pi) (P_+\ox P_-) (\id-\Pi)
  %   = (P_+\ox P_-) - \Pi(P_+\ox P_-)
  %     - (P_+\ox P_-)\Pi + \Pi(P_+\ox P_-) \Pi,\\
  % \end{equation}
  or, equivalently,
  \begin{equation}\label{eq:Cd_sym_constraint}
    \Pi (P_+\ox P_-) \Pi = \Pi (P_+\ox P_-).
  \end{equation}
  Together, \cref{eq:Cd_constraint,eq:Cd_sym_constraint} imply that if
  \cref{eq:Cd_constraint} is satisfied then $\Pi$ and $P_+\ox P_-$
  commute. Conversely, it is easy to see that \cref{eq:Cd_constraint} is
  satisfied if $\Pi$ commutes with $P_+\ox P_-$. Thus the subspace $S$ is
  in $C_d(\RR,d_A)$ iff $\Pi$ commutes with $(P_+\ox
  P_-)=(\id+X)\ox(\id-X)$ and $i\,\Pi\,i^T = \Pi$.

  We will first consider the $P_+\ox P_-$ and $\flip$ symmetries. Since
  $\Pi$ commutes with $P_+\ox P_-$, it must be of the form $\Pi =
  \Pi_{P_{+-}} + \Pi_{P_{+-}}^\perp$ where $\Pi_{P_{+-}}$ is a projector
  onto a subspace in the support of $P_+\ox P_-$, and
  $\Pi_{P_{+-}}^\perp$ is a projector onto a subspace in the orthogonal
  complement thereof. Note that, as we are working in the real vector
  space, $P_+\ox P_-$ is rank $d_A^2/2$. Now, $\flip$ exchanges $P_+\ox
  P_-$ with $P_-\ox P_+$, so $\flip\,\Pi\,\flip^T = \Pi_{P_{-+}} +
  \Pi_{P_{-+}}^\perp$, where $\Pi_{P_{-+}}$ ($\Pi_{P_{-+}}^\perp$) is a
  projector onto a subspace in the (orthogonal complement of the) support
  of $P_-\ox P_+$. But $\flip\,\Pi\,\flip^T = \Pi$, so $\Pi_{P_{-+}}$
  must commute with $\Pi_{P_{+-}}^\perp$ and, furthermore, $\Pi_{P_{-+}} =
  \flip\,\Pi_{P_{+-}}\flip^T$. Thus
  \begin{equation}
    \Pi = (\Pi_{P_{+-}}+\flip\,\Pi_{P_{+-}}\flip^T) + \Pi^\perp,
  \end{equation}
  where $\Pi^\perp$ is a projector onto a subspace in the support of $\id
  - (P_+\ox P_- + P_-\ox P_+)$ that satisfies $\flip\,\Pi^\perp\,\flip^T
  = \Pi^\perp$. Let $r \leq d$ denote the rank of $\Pi_{P_{+-}}$. Since
  $P_+\ox P_-$ has $d_A^2/2$ dimensional support, $r$ cannot be larger
  than this. Also, as $\Pi$ has rank $2d$, $\Pi^\perp$ has rank $2(d-r)$.
  But $\Pi^\perp$ must live in the support of $\id - (P_+\ox P_- + P_-\ox
  P_+)$ which has dimension $d_A^2$, so we require $2(d-r) \leq d_A^2$.
  Thus $r$ is constrained to take values in the range
  \begin{equation}\label{eq:r}
    \max\left[0,d-d_A^2/2\right] \leq r \leq \min\left[d,d_A^2/2\right].
  \end{equation}

  Now consider the $i$ and $\flip$ symmetries. Since $P_+\ox P_- + P_-\ox
  P_+$ is invariant under both these operations,
  $\Pi_{P_{+-}}+\flip\,\Pi_{P_{+-}}\flip^T$ and $\Pi^\perp$ must satisfy
  these symmetries independently. We first focus on $\Pi^\perp$. Let
  $\flip_\pm$ denote the $\pm 1$ eigenspaces of $\flip$. Since
  $\Pi^\perp$ commutes with $\flip$, it must be the sum of a projector
  onto a subspace of $\flip_+$ and a projector onto a subspace of
  $\flip_-$. In other words, $\Pi^\perp = \Pi^\perp_+ + \Pi^\perp_-$
  where $\Pi_{\pm}\,\flip = \flip\,\Pi_{\pm} = \pm \Pi_{\pm}$. Since $i$
  and $\flip$ anti-commute, $i$ must map $\flip_\pm$ to $\flip_{\mp}$.
  Thus $i\,\Pi^\perp_\pm i^T$ is a projector onto $\flip_\mp$. Combined
  with the fact that $i\,\Pi^\perp i^T = \Pi^\perp$ we obtain
  $i\,\Pi^\perp_\pm i^T = \Pi^\perp_\mp$. We can thus assume that
  \begin{equation}
    \Pi^\perp = \Pi^\perp_+ + i\,\Pi^\perp_+ i^T
  \end{equation}
  where $\Pi^\perp_+$ is a projector onto $\flip_+$ within the support of
  $\id - (P_+\ox P_- + P_-\ox P_+)$. Since $\Pi^\perp$ has rank $2(d-r)$,
  $\Pi^\perp_+$ must have rank $d-r$.

  Turning now to $\Pi_{P_{+-}}+\flip\,\Pi_{P_{+-}}\flip^T$, this already
  commutes with $\flip$, so we must be able to rewrite it as
  $\Pi_{P_{+-}}+\flip\,\Pi_{P_{+-}}\flip^T = \Pi_+ + \Pi_-$ where
  $\Pi_{\pm}$ are projectors onto $\flip_\pm$ within the support of
  $P_+\ox P_- + P_-\ox P_+$. By the same argument as before, the $i$
  symmetry imposes $\Pi_- = i\,\Pi_+i^T$, so
  \begin{equation}
    \Pi_{P_{+-}}+\flip\,\Pi_{P_{+-}}\flip^T = \Pi_+ + i\,\Pi_+i^T.
  \end{equation}
  Since $\Pi_{P_{+-}}+\flip\,\Pi_{P_{+-}}\flip^T$ has rank $2r$, $\Pi_+$
  must have rank $r$.

  Finally, consider the $X \ox X$ symmetry. Since $X\otimes X$ commutes
  with $\flip$ and $P_{+-}\ox P_{-+}$, we have that $\Pi_+$ and
  $\Pi^\perp_+$ must also commute with $X \ox X$. This means we can write
  $\Pi_+$ as $\Pi_{++} + \Pi_{+-}$ and $\Pi^\perp_+$ as $\Pi^\perp_{++} +
  \Pi^\perp_{+-}$, where $\Pi_{+\pm},\Pi^\perp_{+\pm}$ are projectors
  onto subspaces of the $\pm 1$ eigenspace of $X\ox X$. Since $\Pi_+$ has
  rank $r$, the ranks of $\Pi_{++}$ and $\Pi_{+-}$ must sum to
  $r$. Similarly, $\Pi^\perp_+$ has rank $d-r$, so the ranks of
  $\Pi^\perp_{++}$ and $\Pi^\perp_{+-}$ must sum to $d-r$. Thus we have
  finally that
  \begin{multline}
    \Pi = \Pi_{++} + \Pi_{+-} + \Pi^\perp_{++} + \Pi^\perp_{+-}\\
          +\; i(\Pi_{++} + \Pi_{+-} + \Pi^\perp_{++} + \Pi^\perp_{+-})i^T.
  \end{multline}

  Conversely, if $\Pi_{++}$, $\Pi_{+-}$, $\Pi^\perp_{++}$ and
  $\Pi^\perp_{+-}$ are arbitrary projectors with the appropriate supports
  and with ranks summing to $r$ and $d-r$, respectively, then a $\Pi$ of
  the above form projects onto a subspace in $F_d(\RR,d_A)\cap
  C_d(\RR,d_A)$. For each value of $r$ satisfying \cref{eq:r}, if
  $\Pi_{++}$ and $\Pi^\perp_{++}$ have ranks $k_1$ and $k_2$, then our
  choice of $\Pi$ is equivalent to choosing an element of
  $\gr_{k_1}(\RR^{d_A^2/2}) \times \gr_{r-k_1}(\RR^{d_A^2/2}) \times
  \gr_{k_2}(\RR^{d_A^2/2}) \times \gr_{d-r-k_2}(\RR^{d_A^2/2})$.
\end{IEEEproof}

This structure lemma allows us to deal with the remaining conditions of
\cref{thm:S_strongly_unextendible_conditions}, namely
\cref{eq:S_positivity,eq:Sperp_positivity}, using probabilistic
arguments.
\begin{theorem}\label{thm:symmetric_positive-semidefinite}
  If $d_A$ is even, and $\lfloor d/2 \rfloor \leq d_A^2/2 - 2$, then the
  set
  \begin{equation}
    \begin{split}
      &P_d(d_A)\\
      &\mspace{10mu}
        =\{S\in F_d(\CC,d_A)\cap C_d(\CC,d_A)\\
      &\mspace{60mu} |\,
        \exists M \in \mat(S), M'\in \mat(\id\ox X\cdot S^\perp):
        M,M' \geq 0
      \}
    \end{split}
  \end{equation}
  has non-zero measure in $F_d(\CC,d_A)\cap C_d(\CC, d_A)$.
\end{theorem}
\begin{IEEEproof}
  Since $\dim\gr_k(\RR^{d_A^2/2}) = (d_A^2/2 - k)k$, we have
  \begin{align}
      &\dim\Bigl(
        \gr_{k_1}(\RR^{d_A^2/2}) \times \gr_{r-k_1}(\RR^{d_A^2/2})\notag\\
        &\mspace{80mu}\times \gr_{k_2}(\RR^{d_A^2/2}) \times
         \gr_{d-r-k_2}(\RR^{d_A^2/2})
     \biggr)\notag\\
      &\mspace{10mu}
      \begin{aligned}
          &=\left(\frac{d_A^2}{2} - k_1\right)k_1
          \left(\frac{d_A^2}{2} - r + k_1\right)(r-k_1)\;\\
          &\mspace{50mu}
          \cdot\left(\frac{d_A^2}{2} - k_2\right)k_2
          \left(\frac{d_A^2}{2} - d + r + k_1\right)(d - r - k_2),
      \end{aligned}
  \end{align}
  which takes its maximum value at $r=d/2$, $k_1=k_2=d/4$ for $d$ a
  multiple of~4, or the closest integers to this otherwise. This means
  that all but a measure-zero subset of $F_d(\CC,d_A)\cap C_d(\CC,d_A)$
  is contained in the component associated with these values of $r$,
  $k_1$ and $k_2$. Indeed, if $d$ is a multiple of~4 then the component
  of $F_d(\CC,d_A)\cap C_d(\CC,d_A)$ corresponding to
  $\gr_{d/4}(\RR^{d_A^2/2}) \times \gr_{d/4}(\RR^{d_A^2/2}) \times
  \gr_{d/4}(\RR^{d_A^2/2}) \times \gr_{d/4}(\RR^{d_A^2/2})$ has measure~1
  in $F_d(\CC,d_A)$. Otherwise, the components corresponding to the
  closest integers to $r=d/2$, $k_1=k_2=d/4$ together have total
  measure~1, with the measure split equally between them. For the
  remainder of the proof we will take $r=d/2$, $k_1=k_2=d/4$ ($d$
  divisible by~4) or any set of closest integers to these. Let
  $K_d(\CC,d_A)$ denote the corresponding part of $F(\CC,d_A)\cap
  C_d(\CC,d_A)$.

  It suffices to show that $P_d(d_A) \cap K_d(\CC,d_A)$ has positive
  measure in $K_d(\CC,d_A)$. To do so, we first construct a subspace
  $S\in K_d(\CC,d_A)$ that contains a positive-\emph{definite} element
  (i.e.\ $M > 0$ for some $M \in \mat(S)$), such that $(\id\otimes
  X)S^\perp$ \emph{also} contains a positive-definite element. This will
  guarantee that every $S'\in K_d(\CC,d_A)$ that is sufficiently close to
  $S$ will contain a positive-semidefinite element, hence will belong to
  $P_d(d_A)\cap K_d(\CC,d_A)$, implying that this set has non-zero
  measure and proving the theorem.

  To construct the desired $S$, choose $S$ to contain $\ket{\omega} =
  \sum_{i=1}^{d_A}\ket{i,i}$, which has $M = \matnorm(\ket{\omega}) = \id
  \geq 0$. We will also require that $S$ be orthogonal to $(\id\ox
  X)\ket{\omega}$ so that $(\id\ox X)S^\perp$ also contains
  $\ket{\omega}$ and is positive definite. (Note that this only works if
  $d_A$ is even, otherwise $\ket{\omega}$ and $(\id\ox X)\ket{\omega}$
  are not orthogonal.)

  $P_\pm\ox P_\mp\ket{\omega} = P_\pm\ox P_\mp (\id\ox X)\ket{\omega} =
  0$, so both $\ket{\omega}$ and $(\id\ox X)\ket{\omega}$ are contained
  in the support of $\id - (P_+\ox P_- + P_-\ox P_+)$. They also both
  belong to the $+1$ eigenspace of $X \otimes X$. Thus to choose $S$ we
  need only choose an additional $k_2-1$ dimensions for $\Pi^\perp_{++}$
  (from a space of dimension $d^2/4-1$) as well as an arbitrary
  rank-$(d-r-k_2)$ projector $\Pi^\perp_{+-}$ whose support is contained
  within the portion of the $-1$ eigenspace of $X\otimes X$ orthogonal to
  $\Pi_\pm\ox P_\mp$ (also of dimension $d_A^2/4$), and arbitrary
  rank-$k_1$ and $r-k_1$ projectors $\Pi_{++}$ and $\Pi_{+-}$. This is
  possible as long as $k_2\leq d_A^2/4$, $d-r-k_2 \leq d_A^2/4$, $k_1
  \leq d_A^2/4$ and $r-k_1 \leq s_A^2/4$. Substituting our choice of $r$,
  $k_1$ and $k_2$, we find that it suffices to take $\lceil d/4 \rceil
  \leq d_A^2/4$.
\end{IEEEproof}

\Cref{cor:symmetric_strongly_unextendible} shows that, for suitable
dimensions, a subspace chosen at random subject to the symmetry and
orthogonality constraints of
\cref{eq:S_conjugate_symmetry,eq:Sperp_conjugate_symmetry,eq:S_orthogonality}
from \cref{thm:S_strongly_unextendible_conditions} will, with probability
1, satisfy the strong unextendibility conditions of
\cref{eq:S_strongly_unextendible,eq:Sperp_strongly_unextendible}. But
\cref{thm:symmetric_positive-semidefinite} shows that there is a non-zero
probability that such a random subspace will satisfy the positivity
conditions of \cref{eq:S_positivity,eq:Sperp_positivity}. Therefore, for
suitable dimensions, there must exist at least one subspace $S$
satisfying all the conditions of
\cref{thm:S_strongly_unextendible_conditions}. Hence, by that theorem,
there exists a pair of channels $\cE_{1,2}$ with $C_0(\cE_{1,2})=0$ but
$Q_0(\cE_1\ox\cE_2) \geq 1$.

Satisfying all the dimension requirements of
\cref{cor:symmetric_strongly_unextendible,thm:symmetric_positive-semidefinite}
imposes constraints on the channel input and output dimensions $d_A$ and
$d_B$, and number of Kraus operators $d_E$ (which corresponds to the
subspace dimension $d$). Together, these constraints impose $d_A \geq 48$
and $d_E \geq 12(2d_A-1)$, giving our main result:
\begin{theorem}\label{thm:main}
  Let $d_A=48$, $d_E = 12(2d_A-1)=1140$ and $d_B = d_Ad_E = 54720$. Then
  there exist channels $\cE_1,\cE_2$ such that:
  \begin{itemize}
  \item Each channel $\cE_{1,2}$ maps $\CC^{d_A}$ to $\CC^{d_B}$ and has
    $d_E$ Kraus operators.
  \item Each channel $\cE_{1,2}$ has no \emph{classical} zero-error
    capacity (hence no quantum zero-error capacity either).
  \item The joint channel $\cE_1\otimes\cE_2$ has positive \emph{quantum}
    zero-error capacity (hence \emph{all} other capacities are non-zero).
  \end{itemize}
\end{theorem}
This trivially implies that there exist channels with similar properties
in all dimensions larger than these, too.

\section{Conclusions}\label{sec:conclusions}
There has been a recent a surge of progress in the theory of quantum
channels, especially their capacities. We now know that two uses of a
quantum channel can sometimes, by using entangled signal states, transmit
more than twice as much classical information as a single
use~\cite{Hastings}. This makes it likely that any expression for the
classical capacity will require regularisation, implying that it cannot
be computed in general. We have known for some time that this is also the
case for the quantum capacity~\cite{DSS97}, but we now also know that the
quantum capacity itself is non-additive. Indeed, it exhibits the
particularly extreme form of non-additivity known as superactivation
\cite{graeme+jon}. This implies that the amount of quantum information
that can be sent through a channel depends on what other channels are
also available. Understanding these additivity violations is now a key
goal of quantum information theory.

Both manifestations of non-additivity---regularisation and non-additive
capacity---are already displayed by the \emph{zero-error} capacity of
\emph{classical} channels~\cite{Shannon_zero-error,Haemers79,alon-union},
though superactivation remains impossible even in the zero-error setting.
Zero-error capacities have been the subject of intense study in the
classical information theory literature for over half a century. They are
therefore an interesting area in which to probe \emph{quantum} channel
capacities, and attempt to understand non-additivity phenomena.
Non-additivity in the purely classical setting obviously has nothing to
do with entanglement. But quantum channels display even stronger
non-additivity than their classical counterparts. In the quantum world,
the presence of entanglement \emph{does} lead to superactivation of the
classical zero-error capacity of quantum channels~\cite{CCH09}.

The usual classical and quantum capacities are not at all closely
related. There is no reason to expect that channels displaying additivity
violations for the quantum capacity will possess any interesting
additivity properties for the classical capacities, or vice versa. As a
consequence, the recent non-additivity results for the usual
capacities~\cite{graeme+jon,Hastings} required very different
mathematical techniques for the two cases.

However, in the zero-error setting, this work shows a striking
non-additiv\-ity phenomenon that connects the classical and quantum
capacities. We have proven the existence of pairs of channels that,
individually, can not communicate \emph{any} information with zero error,
even classical information. But, when used together, even a single use of
the joint channel suffices to communicate \emph{all} forms of
information, quantum and classical. These channels therefore exhibit the
most extreme possible form of additivity violation; their zero-error
capacities \emph{simultaneously} violate additivity for both classical
and quantum information, and in the most extreme way (superactivation) to
boot. This extreme form of superactivation is trivially impossible for
classical channels, or for the usual capacities of quantum channels.
Zero-error communication therefore provides a compelling setting in which
to explore non-additivity phenomena in quantum information theory.

\bibliographystyle{IEEEtran}
\bibliography{SuperDuper}

\end{document}